\documentclass[fleqn,twoside]{article}
\usepackage{espcrc2}

\usepackage{graphicx}

\newcommand{\AmS}{{\protect\the\textfont2
     A\kern-.1667em\lower.5ex\hbox{M}\kern-.125emS}}
\newcommand{\as}{\alpha_s}
\newcommand{\bea}{\begin{eqnarray}}
\newcommand{\eea}{\end{eqnarray}}
\newcommand{\be}{\begin{equation}}
\newcommand{\ee}{\end{equation}}
\newcommand{\pt}{{\rm PT}}
\newcommand{\eps}{\varepsilon}

\title{Some Recent Developments in Perturbative Quantum Chromodynamics\thanks{YITP-SB-02-02,
based on a talk at {\it
Light-Cone Physics, Particles and Strings, Trento, September\ 2001.}}}

\author{George Sterman\address[MCSD]{C.N.\ Yang Institute for
Theoretical Physics\\
           State University of New York at Stony Brook, Stony Brook, NY
11794-3840, USA}%
          }

\begin{document}

\begin{abstract}
This talk reviews some recent trends in perturbative quantum chromodynamics,
with emphasis on higher orders in perturbation theory,
resummation and power corrections.
\vspace{1pc}
\end{abstract}

\maketitle

\section{INTRODUCTION}

This talk begins with a few general comments about
the place of QCD studies.  I go on to give
a few words on the central role played
by factorization and related concepts.
A brief mention of some important steps forward in
multiloop calculations is followed by a description
of progress in abstracting classes of corrections
at arbitrary order in perturbation theory.  
I will argue that the interplay of resummation with
power corrections is a link between perturbation
theory and the nonperturbative degrees of freedom
of QCD.

\begin{figure}[h]
\hbox{\hskip 0.8 true cm
\includegraphics[width=15pc]{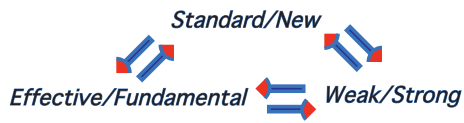}}
\caption{Dual concepts.}
\label{dual}
\end{figure}

A preliminary message in this talk
is that QCD should be thought of as the exemplary
field theory, one which illustrates all of
the paradigms of what might be
called ``postmodern" particle physics, characterized
by pairs of complementary concepts, as illustrated
in Fig.\ \ref{dual}.

QCD is the meeting ground of all the interactions
of the standard model, and indeed, although it
has only a single dimensional scale itself, $\Lambda_{\rm QCD}$,
the influence of electroweak scales, through the W, Z and
quark masses, is profound.  At the same time, QCD
generates its own secondary scales, though chiral
symmetry breaking, confinement and vacuum structure.
In so doing, it provides a tapestry of effective
theories, each  fitting into an
energy range appropriate to a particular set of
states.  Among this list are perturbative QCD, heavy
quark effective theory, non-relativistic QCD, lattice
QCD, and nuclear physics.  There is no doubt that
QCD is ``correct", at least the way classical
electricity and magnetism is correct, as itself an
effective theory appropriate to a wide range of
phenomena.  Tests of QCD are, in this sense, tests
of quantum field theory itself.

\begin{figure}[h]
\includegraphics[width=23pc]{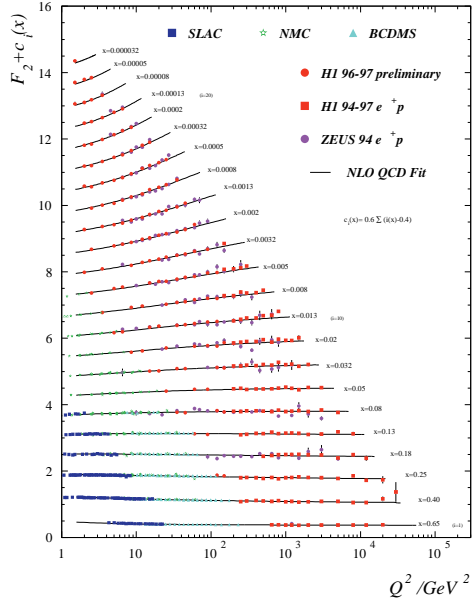}
\label{F2}
\caption{$F_2$ data}
\end{figure}

\begin{figure}[h]
\includegraphics[width=15pc]{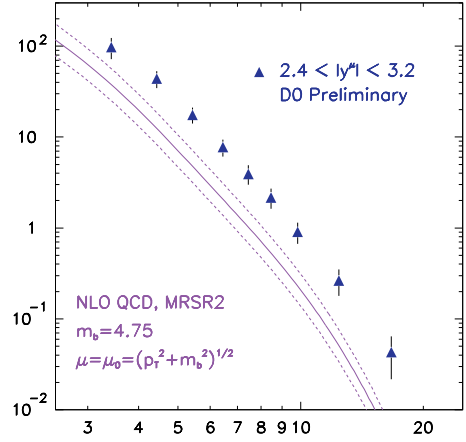}
\caption{b quark data}
\label{bqpt}
\end{figure}

We are in a ``golden age" for hadronic data,
both in coverage and
quality.   The great accelerators of the nineteen nineties,
HERA, the Tevatron, SLC and LEP,
brought about a veritable revolution in strong interaction
data.  Our standards for judging success
in fits to this data have correspondingly tightened.
In this light, we can judge the experimental and
theoretical successes and limitations of our description of
deep-inelastic structure functions, Fig.\ 2,
side-by-side with data on the transverse momentum
distribution of b quarks, Fig.\ \ref{bqpt}.

\section{UNITY OF QCD FACTORIZATIONS}

The basis of perturbative QCD (PQCD) is
in asymptotic freedom, applied to quantities that
are {\it infrared safe}, that is, which depend
only upon one (or more) momentum scales
much larger than $\Lambda_{\rm QCD}$, up to
corrections that can be estimated.
For the simplest cases, observables like
$\rm e^+e^-$ total and jet cross sections,
it is convenient to
choose the (renormalization) scale of
the QCD coupling to equal the hard scale, $Q$,
\begin{eqnarray}
Q^2\; \hat \sigma(Q^2,\mu^2,\alpha_s(\mu))
&\ & \nonumber\\
&\ & \hspace{-30mm}
= \sum_n c_n(Q^2/\mu^2)\; \as^n(\mu) + {\cal O}\left({1\over Q^p}\right)
\nonumber\\
&\ & \hspace{-30mm} = \sum_n c_n(1)\; \as^n(Q) +   {\cal
O}\left({1\over Q^p}\right)
\, .
\label{irs}
\end{eqnarray}
In the second form, the cross section is an
expansion in a small parameter $\alpha_s(Q)$, with
dimensionless coefficients.  Corrections
are suppressed by powers of the large scale.
There are two challenges in treating infrared
safe cross sections:  the calculation of
the $c_n$, and the interpretation and estimation
of power corrections.

The set of infrared safe observables is rather
limited.  The applicability of perturbative methods
is  greatly expanded by {\it factorization},
which can be applied whenever an observable can
be written as a product, or a convolution (denoted $\otimes$), of a
short-distance function, times one or more functions
that absorb nonperturbative, long-distance dynamics,
\bea
Q^2\sigma(Q,m)
&=&
\omega_{\rm SD}(Q/\mu,\as(\mu))\otimes f_{\rm LD}(\mu,m)
\nonumber\\
&\ & + {\cal O}\left({1\over Q^p}\right)\, ,
\label{fact}
\eea
where $\mu$ is the factorization scale, which separates
the short- and long-distance components.  This sort of
factorization is familiar from deep-inelastic scattering (DIS)
inclusive cross sections,
where the $f$'s are parton distributions (PDFs),
and from hard scattering cross sections at hadronic
colliders, but it also applies, in varying forms, to
amplitudes for exclusive processes,
such as B decays and deeply-virtual Compton scattering,
It also applies  to jet cross sections in $\rm e^+e^-$ annihilation,
for events where the jets are very narrow, and where
the cross section is sensitive to relatively long-time behavior.

Different processes demand different proofs 
for their factorizations,
but there are a number of common themes.  At the
basis of all factorization is the quantum-mechanical incoherence
between short- with long-distance dynamics, and also between different
sources of long-distance dynamics that
develop at spacelike separations.  The first
allows the operator product expansion in DIS,
the second justifies factorization for hadronic collisions.

The proverbial ``new physics" is to be found in $\omega_{\rm SD}$;
the long-distance $f_{\rm LD}$'s are  ``universal", simply
in the sense that they do not depend on what's going on
at short distances.
Factorization may be regarded from
complementary viewpoints.  On the one hand, we may think of
$\omega_{\rm SD}$
as the perturbation expansion in
a theory with $f_{\rm LD}$ an IR regulator.
This would be a ``Wilsonian" point of view, with $\mu$ as
an infrared cutoff.  Another view is to
describe $f_{\rm LD}$ as a matrix element in an effective theory with
$\hat \sigma_{\rm SD}$ a
matching coefficient, and $\mu$ the renormalization scale.

However we describe factorization, it can be useful
only when the factorization scale $\mu$ is large enough
that we can use perturbation theory to compute
the short-distance function.
A consequence of this rather simple observation is that,
whenever there is factorization, there is
a calculable evolution, which can be derived
from the independence of the physical cross section
from the factorization scale.  Denoting infrared mass scales by $m$,
we have
\bea
0=\mu{d\over d\mu} \ln \sigma_{\rm phys}(Q,m)
&\ & \nonumber\\
&\ & \hspace{-40mm} =
\mu{d\over d\mu} \ln\left\{\omega_{\rm SD}(Q/\mu,\as(\mu))\otimes
f_{\rm LD}(\mu,m) \right\}\, ,
\eea
so that, very schematically,
\be
\mu{d \ln f\over d\mu}= - P(\as(\mu)) = - \mu{d \ln \omega \over d\mu}\, ,
\label{evol}
\ee
where the (splitting) function $P(\alpha_s)$ can depend
only on the variables held in common between $\omega_{\rm SD}$
and $f_{\rm LD}$, which are the coupling and
an appropriate convolution
variable (or moment variable).
For factorizations like
Eq.\ (\ref{fact}), where corrections
are power-suppressed, all orders in $\as$ respect the form of the evolution equation.
The solution to any evolution equation 
can be thought of as a resummation of the theory,
\be
\ln \sigma_{\rm phys}(Q,m) \sim
\exp\left\{  -\int^Q {d\mu'\over \mu'} P\left( \alpha_s(\mu')\right)
\right\}\, .
\ee
Let's first discuss examples of
progress in the calculations of the short-distance
functions, and then return to examples of what we have learned
over the past few years from resummations.

\section{HIGHER LOOPS \& JETS}

\subsection{Toward a Two-loop Phenomenology}

\begin{figure}[h]
\includegraphics[width=13pc]{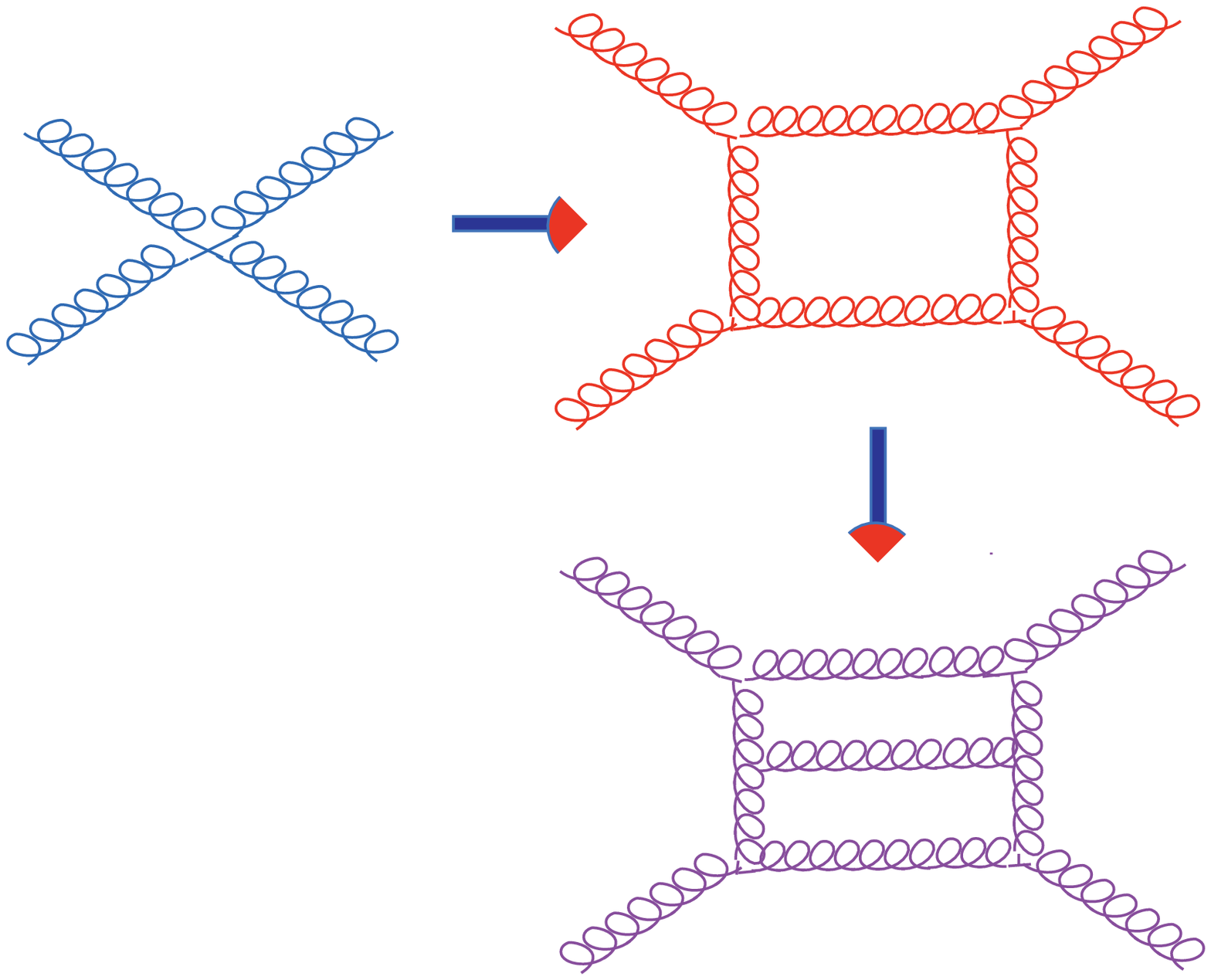}
\end{figure}

The perturbative running coupling is typically quoted
at the scale of the Z mass,
   $\alpha_s(M_Z) \sim 0.12$.  In a rough
sense, then, we can estimate that
corrections at $\alpha_s^2$ are of order 1\%.
Now it so happens that one percent accuracy
in jet cross sections at a linear collider
would allow a much more accurate extrapolation
of Standard Model scales through their logarithmic
evolution to energies characteristic of
grand unification.  At the same time, one percent
is roughly the level to which nonperturbative corrections
have shrunk for inclusive jet cross sections at
a few hundreds of GeV.  The scales of the physics
in which we are most interested strongly invite
us to two loops.

Referring to Eq.\ (\ref{irs}), a summary of the
status of $c_n$'s is,  they are known at
next-to-leading order
(NLO)
   up to four jets $\rm e^+e^-$, and up to
NNLO only for one-scale problems: the $\rm e^+e^-$ total
cross section and DIS structure functions.
Two loops is the frontier of finite-order PQCD.
The splitting function for DIS factorization
(the DGLAP kernels),  $P(z)$
are themselves known at order $\alpha_s^2$,
although to make full use of two-loop calculations
of short-distance functions, we should know them at
three loops.  Here, progress has been made,
and an increasing number of moments $\int dz z^nP(z)$,
at $\alpha_s^3$ have been reported
\cite{Lar97,Moc01}.
Using these results, there is a
budding phenomenology of DIS at NNLO 
\cite{nnlo}.

In very recent and substantial progress, two-loop QCD corrections
to $2\rightarrow 2$ quark and gluon scattering amplitudes
have been computed \cite{Ber00,Glo01}.
These results are the first step along
what promises to be a very steep path to
   the computation of cross sections, including phase-space integrals
at NNLO.
In the estimates of the experts,
we are still a number of years away from true NNLO jet cross sections,
but it should be worth the wait.

\subsection{Energy Flow}

One fundamental use of the two-loop scattering
amplitudes that are beginning to appear, and
the corresponding $2\rightarrow 3$ and $2\rightarrow 4$
processes that will follow, is in the computation
of jet production in hadronic
collisions, which will be a background
(and signal) for new physics.  Closely related
calculations will also be applied to event shapes
in $\rm e^+e^-$ annihilation \cite{qqbarg}.
NLO inclusive
jet cross sections are a familiar feature from
the Tevatron.  

For the most part, jet events
are identified ``algorithmically", by a series of steps
that assigns particles (or calorimeter towers)
in an event to a list of jets.  Jet algorithms have
become quite sophisticated, and are becoming more
so  \cite{Elli01}.  
Newer versions avoid some of the technical problems
that have been associated with previous implementations.
The primary requirement of a jet algorithm is infrared
safety, that is, it should specify cross
sections that are factorizable into PDFs and
a pertrubative short distance function.

Here, I would like to mention a class of observables that are supplements
to jet algorithms, and 
which quantify the flow of energy  \cite{Tka94,Gie97}.
One of the first, and still familiar, infrared safe
cross sections in electron-positron annihilation
is the energy-energy correlation \cite{eec}.  Energy flow
variables are closely related to this concept.
These observables are weighted sums that
are linear in the momenta of final-state particles.
In a notation\footnote{Described to me by Walter Giele.} 
based on the formalism of \cite{Tka94}, they
may be represented conveniently as
\bea
J_\mu(m\equiv \{\eta,\phi\})
&\ & \nonumber\\
&\ & \hspace{-20mm} = \int dm'\, P_\mu(m')\, \times {\cal W}(m'-m)\, ,
\eea
where the integral is over directions $m'$ and ${\cal W}$ is the weighting
function, and where 
\bea
P_\mu(m') &\equiv& \sum_{i=1}^N\; p_\mu^i\, \delta\left(m'-m_i\right)\, ,
\eea
for a state with $N$ final state particles.
Properly chosen, the weight function themselves can
distinguish the gross features of jet events \cite{Tka94,Gie97}.  They may  also be readily
treated with the techniques of resummation and,
eventually, estimates of power corrections,
about which we shall say a bit more below.

A feature of jet cross sections is that
they discard much of event structure.
It may be valuable to measure a variety
of weighted energy flow correlations for events
with large transverse momenta,
to derive information
that is complementary to differential jet
cross sections in terms of transverse momentum and rapidity.
For example, correlations involving relatively soft
particles may well contain information on
the flow of color at short distances.

In time,
our understanding of QCD will progress, and we would
like to be able to go back to the data to ask new
questions.  Certainly, an ideal to be sought after is to
free the data from contemporary theory.  This is a
difficult task, and complete archival accessibility
may be an unattainable goal.  Nevertheless, it is
worthwhile to try to keep the reported data as
close as possible to the observables themselves,
rather than to report on secondary quantities.
Thus, in DIS, it is important
 to make available measured cross sections,
in addition to structure functions, and certainly
to fitted parton distributions.

\section{THRESHOLD RESUMMATION}

Now let us turn to an example that shows
how it is sometimes possible to organize and compute
a class of corrections to all orders in
perturbation theory, in this case
through threshold resummation \cite{Kido99}.

\subsection{Refactorization and Evolution}

We consider a short-distance cross section
to produce a final state F, of total
invariant mass $Q$.
Examples include Z production,
with $Q=m_{\rm Z}$, and the production of
a pair of jets of total invariant mass $m_{\rm JJ}=Q$,
in p$\rm \bar p$ collisions.
The basic factorization theorem is
\bea
Q^2\sigma_{AB\rightarrow F}(Q)
&=&
   {f_{a/A}(x_a,\mu)} \otimes  {f_{b/B}(x_b,\mu)}
\nonumber\\
&\ & \hspace{-15mm}
\otimes\,
   {\omega_{ab\to
F}\left(z={Q^2\over x_ax_b S},Q,\mu\right)}\, .
\label{abfact}
\eea
In the short distance function $\omega$,
the limit $z\rightarrow 1$, is special.  At this
point, the partons have only just enough energy
to produce the observed system F, with nothing
left over for radiation.  Not surprisingly,
the short-distance function is singular at $z=1$,
although the singularities are integrable.
Their integrability is guaranteed by the
factorization theorem itself.  At order $r$,
we encounter terms,
\bea
\omega_{ab\to F}^{(r)}(z) &\sim& \left({\alpha_s\over
\pi}\right)^r
{1\over r!}
\; \left[{\ln^{2r-1}(1-z)\over 1-z}\right]_+\, ,
\label{plusdist}
\eea
with ``plus" distributions", defined by their
integrals with smooth functions (in this
case the PDFs): $\int_0^1 dz g(z)[f(z)]_+\equiv
\int_0^1(g(z)-g(1))f(z)$.
In general, the limit $z\rightarrow 1$
corresponds to ``elastic" kinematics, in
which the masses of final state jets
are bounded by $m^2[J_i]\le (1-z)S$.
In this limit, bremsstrahlung gluons radiated
by the partons involved in the hard scattering
are forced to be collinear to the parent parton, as
pictured in Fig.\ 4.

\begin{figure}[h]
\hbox{\hskip 1.0 true cm
\includegraphics[width=15pc]{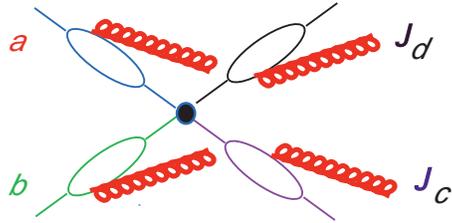}}
\label{elasticlim}
\caption{Collinear radiation in the elastic limit.}
\end{figure}

Near $z=1$, we can invoke the principles of
quantum mechanical incoherence mentioned above: that the evolutions
of the final state jets decouple, not only from the
hard scattering, but also from each other.
This allows us to ``refactorize" the hard scattering
function $\omega$ in Eq.\ (\ref{abfact}) into
convolutions of functions involving 
the initial, and also the final state jets,
as well as soft radiation between the jets.
As announced above, such a refactorization
leads to new equations for evolution, and
to resummation, in this case 
of the singular terms in $1-z$ of Eq.\ (\ref{plusdist}).
The result is most easily exhibited for
Mellin moments of $\omega$, which exponentiate, in terms of an anomalous
dimension $A(\as)$:
\bea
{\ln}\left\{\int_0^1 dz z^{N-1} \omega_{a\bar a\to F}(z,\mu)\right\}
&\ & \nonumber\\
&\ & \hspace{-50mm} =
-2\int_0^1 dz{z^{N-1}-1\over 1-z}
\int_{(1-z)^2Q^2}^{\mu^2}{d m^2\over m^2} A_a(\alpha_s(m))
\nonumber\\
&\ & \hspace{-50mm} \sim  2\int_{Q^2/N^2}^{\mu^2}{d
m^2\over m^2} A_a(\alpha_s(m))
\, \ln\left( {Nm\over Q}\right)\, .
\label{thresum}
\eea
This turns out to produce an enhancement in the cross section,
which can be large in certain circumstances, but which
is often quite modest.  It shows how, even in
the presence of singular corrections such as
(\ref{plusdist}), next-to-leading order
calculations in factorized cross sections can still
give reliable answers.
Another use of threshold resummation is that it
automatically organizes,
and hence reduces, factorization scale dependence from
higher orders.  We can see how this comes about
by recalling that, as in Eq.\ (\ref{evol}) above,
evolution results from the complementary variations
of the hard and soft functions in a factorized cross
section.  Taking the derivative of the resummed
hard function in Eq.\ (\ref{thresum}), we find
\bea
{d\ln \omega_{a\bar a}(z) \over d\ln\mu^2} &\sim& {2A_a(\as)} \ln N  
\nonumber\\ 
&\sim& -
{d\ln [\tilde f_{a/A}\tilde f_{\bar a/B}]\over d\ln\mu^2}\, ,
\label{resderiv}
\eea
in terms of moments of the PDFs, $\tilde f$. By consistency with Eq.\ (\ref{evol}), the
function $A$ that appears in the
resummation formula {\it must} be the same function
that appears in the $1/[1-z]_+$ term in the splitting
functions.  At NLO, the left-hand relation in
Eq.\ (\ref{resderiv}) holds only to order $\alpha_s$, the
first term in the expansion of $\ln\omega$, while
for the resummed cross section it holds to all orders.
  This leads to an 
important reduction in factorization scale dependence compared to NLO
\cite{Ber98,Kid99,Cat99}.

\subsection{Color Mixing}

\begin{figure}[h]
\hbox{\hskip 1 cm
\includegraphics[width=15pc]{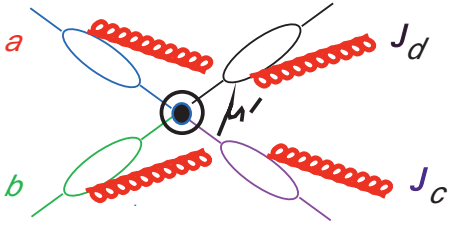}}
\caption{The refactorization scale.}
\label{refactscale}
\end{figure}

The result (\ref{thresum}) shows only the leading
logarithms organized by threshold resummation.
Beyond leading log,
threshold resummation also sheds an interesting light on
the evolution of color from short to long distances.
The evolution equations that lead to the resummation
involve a ``refactorization scale",
$\mu'$, that separates the final state jets
from the hard scattering.  Like the standard
factorization scale, $\mu'$ can vary, in this
case between the jet masses, $m[J_i]$ and the hard scale, $Q$,
as illustrated in Fig.\ \ref{refactscale}.

As $\mu'$ varies, virtual gluons near scale $\mu'$
pass between the hard scattering, and the radiation
from the jets.  The hard scattering, however, involves
color exchange in general, and as gluons move in
and out of the short-distance function, the color
exchange seen at long distances changes.  This
leads to a matrix evolution equation beyond leading
logarithm, which governs how color is radiated from
short to long distances \cite{Kid98}.  From the effective
theory point of view, the hard scattering is
an elementary vertex that requires renormalization,
and the matrix in the evolution equation is 
the anomalous dimension matrix of that vertex.

As an example, we may consider quark pair annihilation into
gluons, a reaction of importance for high-$p_T$ jets
at the Tevatron.  Here a color basis for the reaction
consists of $s$-channel singlet, and two $s$-channel
octets, of the forms $d^{ijk}T_k$ and $f^{ijk}T_k$,
where $T_k$ is a (Gell-Mann matrix) generator of
SU(3) in the fundamental (quark) representation, and
where $f$ and $d$ are the antisymmetric and symmetric
matrices of the SU(3) algebra, respectively.  The $T$'s couple the color
indices of
the incoming quark pair, and $d$ or $f$
those of the outgoing gluons.
The anomalous dimension matrix for this vertex turns
out to be
{\small
\bea
\Gamma_{S'}=\frac{\alpha_s }{\pi} &\ &
\\
&\ & \hspace{-2.0 cm} \times \left(
                  \begin{array}{ccc}
                    0 &   0  & U-T  \vspace{2mm} \\
                    0  &  {C_A\over 2}(U+T) & {C_A\over 2}(U-T)
     \vspace{2mm} \\
             2(U-T) & {N^2-4\over 2N}(U-T)  &  {C_A\over 2}(U+T) \vspace{2mm}
                    \end{array} \right)\, ,
        \nonumber
\eea
}
with $T$ and $U$ defined in terms of the scattering
angle $\theta$, by $T,U\sim \ln(1\mp \cos \theta)$.
The resummation of next-to-leading logarithms is best
carried out by going to a diagonal basis for this
matrix, with a result of the form
\bea
\exp \left\{ - \int_{Q/N}^Q {d m \over m}\; \left[ \lambda_j(\alpha_s(m))
\right]\right\}\, ,
\eea
where $j$ labels an eigenvector of the
color exchange matrix, $\Gamma_S$, and where the $\lambda$'s
are the corresponding eigenvalues.
The same analysis may be carried out for any partonic
reaction, in particular, for
$g+g\rightarrow g+g$, where the coupling of the octet
gluons leads to a 9$\times$9 matrix (which can be reduced
to $8\times$8 \cite{Kid98}).  Precisely the same matrix of anomalous
dimensions appears in the calculation
of the two loop gluon scattering amplitude \cite{Glo01}
mentioned above, where it controls poles in
dimensional regularization of the
$gg\rightarrow gg$ scattering amplitude, through
the function \cite{Cata98}
\bea
I^{(1)}(\epsilon) (s=\mu^2_R)
&\sim&
\left( {1\over \epsilon^2}+{\beta_0\over N_c\epsilon}\right)\, \Gamma_S\, .
\eea
This connection strongly suggests that a combination of resummation
analysis with explicit calculations can lead to further
insights.

\section{POWER CORRECTIONS: \\UNIVERSALITY AND BEYOND}

One feature of threshold resummation that we haven't
yet emphasized is that, for $z$ close enough to one, the $m^2$ integral
on the right-hand side of Eq.\ (\ref{thresum}) goes
through $\Lambda_{\rm QCD}$, where the perturbative
running coupling that appears in function $A$
become undefined.  At $m=\Lambda_{\rm QCD}$, the
perturbative expression
$\alpha_s(m)\sim 2\pi/b_0\ln(m/\Lambda_{\rm QCD})$ actually
has a pole, commonly referred to as its ``Landau pole",
by analogy to the corresponding pole in the QED
running coupling
at very large momentum.  At this scale,
perturbation theory is not self-consistent,
but can be used to identify
nonperturbative
corrections that restore self-consistency.  In this talk, of course, I cannot
do justice to the various approaches to the subject,
but I'll try to give something of the spirit.

\subsection{Resummed Thrust}

\begin{figure}[h]
\hbox{\hskip 1.5 cm
\includegraphics[width=10pc]{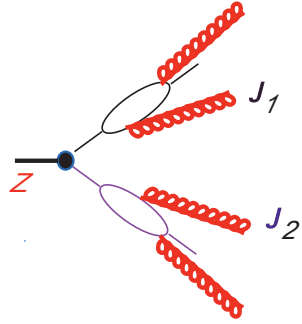}}
\caption{Z decay into two jets.}
\label{Tto1}
\end{figure}

The approach I'll take starts with a resummed cross section for
$\rm e^+e^-\rightarrow Z^*(Q)\rightarrow F$. 
In the ``elastic" limit for $F$, with
two jets of invariant masses $m_i\ll Q$, we have
\bea
1-T \sim (m_1^2+m_2^2)/Q^2 \rightarrow 0\, ,
\eea
where $T$ is the thrust.
The configuration is illustrated in Fig.\ \ref{Tto1}.

As in our discussion of threshold resummation,
light-like relative velocities between the jets
imply factorization.  This leads to evolution and thus to
a cross section resummed in logarithms of $t\equiv 1-T$, 
\bea
{1\over \sigma_{\rm tot}}\, {d\sigma_{\rm PT}(1-T)\over dT}
&\sim&
\label{Tresum}\\
&\ & \hspace{-30mm}
    {d\over dT}\;
\exp\;
\left[- \int_{t}^1 {dy\over y} \int_{y^2Q^2}^{y Q^2}
{d m^2\over m^2}\; A(\as(m))\right]\, ,
\nonumber
\eea
where $A(\as)$ is the same function as 
in threshold resummation.

\subsection{From the Resummed Cross Section \\to Power Corrections}

In the resummed cross section of (\ref{Tresum}),
the Landau pole appears in the limit $y\rightarrow 0$.
 The following form generalizes the leading logarithm
result:
\bea
  \int_0^1 dy{{\rm e}^{y/t}-1 \over y} \int_{y^2Q^2}^{y Q^2}
{d m^2\over m^2}\; A_q(\as(m)) &\ &
\nonumber\\
&\ & \nonumber\\
&\ & \hspace{-50mm}\sim  {{2\over tQ}\int_0^{Q^2} dm\; A_q(\as(m))}
\label{Texpand}\\
&\ & \nonumber\\
&\ & \hspace{-60mm}
+ {\sum_{n=2}^\infty {1\over n n!} 
\int_0^{Q^2} {dm^2 \over m^2} \left({m\over tQ}\right)^{n}\; A_q(\as(m))}
\nonumber\\
&\ & \nonumber\\
&\ & \hspace{-63mm}
- {\sum_{n=2}^\infty {1\over n n!} 
\int_0^{Q^2} {dm^2 \over m^2} \left({m^2\over tQ^2}\right)^{n}\; 
A_q(\as(m))}\, .
\label{expand}
\nonumber
\eea
In this expanded form, the range of integration, $m\sim \Lambda_{\rm QCD}$,
of each term is explicitly suppressed by a power of $Q$.  The first
term in the expansion is the only one that has the power $Q^{-1}$.
 From this point of view, it is natural to think of the
$1/Q$ term as not only dominant, but also 
as reflecting universal properties of the strong coupling.
Similar resummations can be carried out
for a fairly large class of event shapes $e=1-T,B_T,C \dots$,
all constructed to vanish in
the limit of narrow jets.

As a minimal approach, we consider only the first term in the expansion (\ref{Texpand}),
noting that the 
coefficients of the integral over the running coupling
may be different in other event shapes.
Taking this into account, we define \cite{Dok98},
\be
\lambda_e \sim C^{(e)}_{\rm PT}{\cal M}_e\int_0^{\mu_0} dk\, \as(k)\, ,
\label{lambda}
\ee
where we replace the
function $A(\alpha_s)$ by the combination ${\cal M}_e\as$.
The constant ${\cal M}_e$ incorporates certain higher-order
effects that depend on  the choice of $e$.
The integral in (\ref{lambda}) may be considered as a
``universal" nonperturbative parameter.  Its actual value,
of course, depends on  the upper limit,
which serves as a factorization scale  between perturbation
theory and long distance physics.

\begin{figure}[h]
\hbox{\hskip 0.5 cm
\includegraphics[width=15pc]{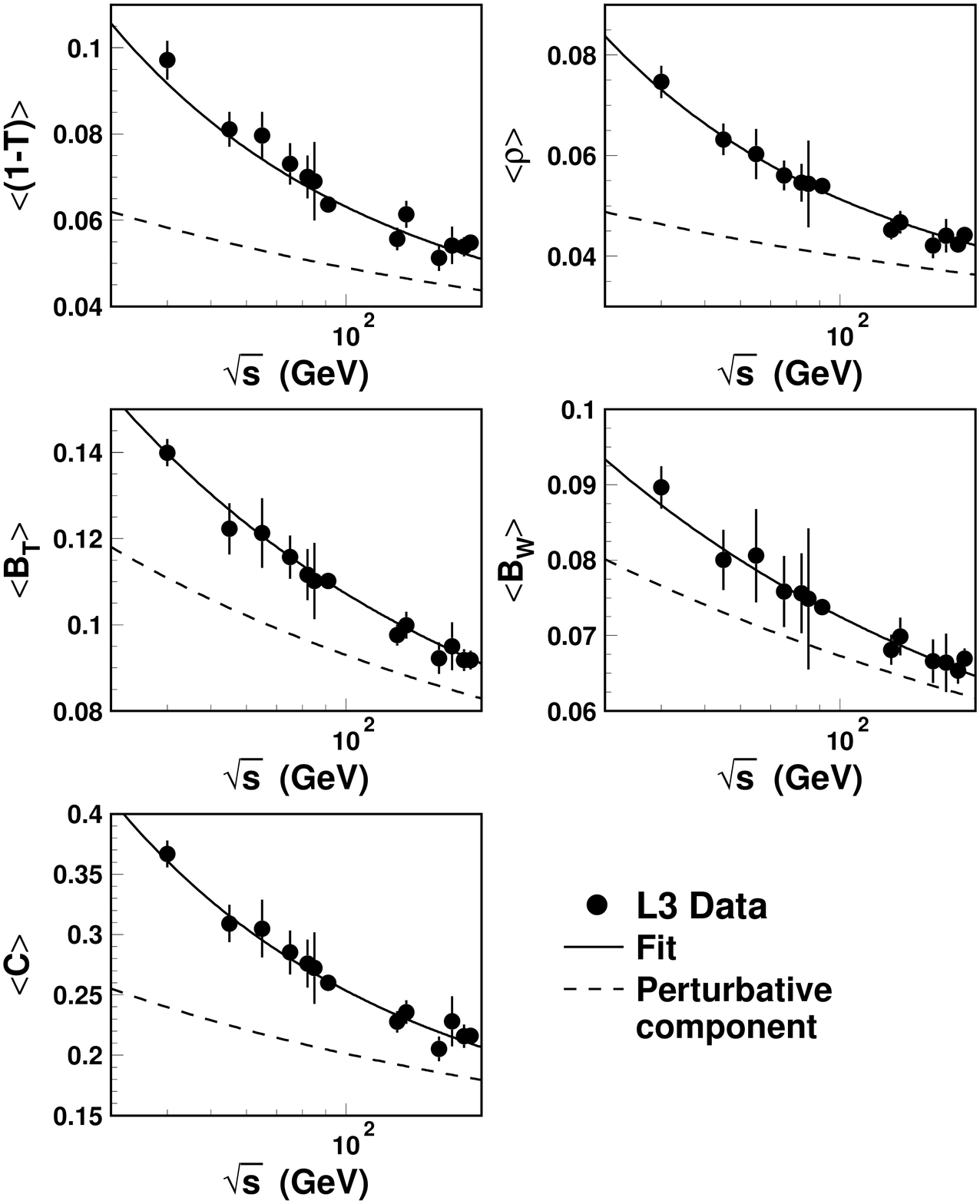}}
\caption{L3 data for first moments of various
event shapes. The dashed line is perturbation
theory, the solid line a fit based on
Eq.\ ({\protect \ref{lambda}})}.
\label{l3}
\end{figure}

This approach predicts 
a simple shift of the cross section \cite{shift},
\bea
{d\sigma(e) \over de} =
{d\sigma_\pt(e-\lambda_e/Q) \over de} +{\cal O}\left({1\over e^2Q^2}\right)\, 
.
\label{shift}
\eea
An interesting consequence 
is that the true $n$th moments of event shape $e$
are related to the perturbative predictions
for the $n$th and $n-1$st moments by
\bea
\langle e^n \rangle = \langle e^n
\rangle_{\rm PT} + n{\lambda_e\over Q}\; \langle e^{n-1}\rangle_{\rm PT}\, .
\eea
Observations of a variety of event shapes
 \cite{L300} fit rather well
with this picture for the first moments, as 
illustrated by Fig.\ \ref{l3}.  At the same time, there 
appear to be
substantial $1/Q^2$
corrections in second moments, such
as $\langle(1- T)^2(Q)\rangle$, and indeed,  purely 
perturbative descriptions can be given for
the first moments \cite{delphi}.

\subsection{Shape Functions}

A natural generalization of (\ref{shift}),
sensitive to corrections beyond the
first term of (\ref{Texpand}) is  a convolution \cite{Kor99},
\bea
{d\sigma \over de} &=&
\int_0^{eQ} d\epsilon\; f_e(\epsilon)\; {d\sigma_\pt(e-\epsilon/Q) \over
de} \nonumber\\
&\ & \hspace{5mm} +{\cal O}\left({1\over eQ^2}\right)\, ,
\label{shapeconvol}
\eea
where the nonperturbative
``shape function" $f_e$ is independent of $Q$.
It is therefore sufficient to fit $f$ at $Q=m_Z$
to derive predictions for all $Q$.  Eq.\ (\ref{shapeconvol})
can incorporate all corrections like $1/(tQ)^n$
in Eq.\ (\ref{expand}).

\begin{figure}[h]
\hbox{\hskip 1.0 cm
\includegraphics[width=15pc]{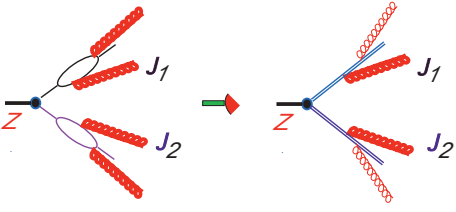}}
\caption{Effective theory for soft radiation in
the elastic limit. The double line represents the
sources of Eq.\ ({\protect \ref{qbarqW}}).}
\label{effective}
\end{figure}

The convolution in (\ref{shapeconvol})
may be based on  effective theory
for soft gluon radiation by $3\otimes 3^*$ sources,
representing a quark pair.  In this theory, radiation 
emerges from a product of ordered exponentials
in the directions of the jets' momenta,
\bea
W(0)&=& P\;{\rm e}^{ig\int_0^\infty d\lambda \beta_1\cdot A(\lambda\beta_1)}\
\nonumber\\
&\ & \times
\left[\, P{\rm e}^{ig\int_0^\infty d\lambda \beta_2\cdot
A(\lambda\beta_2)}\, \right]^\dagger\, ,
\label{qbarqW}
\eea
as shown in Fig.\ \ref{effective}.

In the effective theory,  true universality is at the  level of
correlations of energy flows,
given by the matrix elements
\bea
{\cal G}(\vec n_1\dots \vec n_N)
&=& \langle 0 | W^\dagger(0)
{\cal E}(\vec n_1) \times
\nonumber\\
&\ & \times \dots
\dots {\cal E}(\vec n_N)   W(0) |0\rangle\, ,
\eea
where the operators ${\cal E}$ are defined
to act on states by:
\bea
{\cal E}(\vec n)|N\rangle =
\sum_{i=1}^N\delta^2(\vec n-\vec n_i)\; E_i|N\rangle\, .
\eea
Note the close relationship of these operators
to the jet energy flow operators introduced above.

For event shapes like thrust, which
are related to jet masses,
we need correlations of energy flow into opposite hemispheres,
measured by a shape function of two variables,
\bea
f(\varepsilon_L,\varepsilon_R)
&=&\\
&\ & \hspace{-5mm}
\langle 0|W^\dagger \delta(\varepsilon_L-{\cal E}_L)\, 
\delta(\varepsilon_R-{\cal E}_R)
W|0\rangle\, ,\nonumber
\eea
where we define
\bea
{\cal E}_{L,R} &=& \int d^2\vec n \theta(\pm \cos\theta_n)\,
(1-|\cos\theta_n|)\, {\cal E}(\vec n)\, .\nonumber\\
\eea
In this notation, the squared mass of a jet moving into
the right (left) hemisphere is proportional to $Q\eps_R$ ($Q\eps_L$).
The thrust event shape function is then given by
\bea
f_T(\eps) &=&
\int d\eps_Rd\eps_L\, \delta(\eps-\eps_L-\eps_R)
\nonumber\\
&\ & \hspace{10mm} \times f(\eps_L,\eps_R)\, .
\eea
A ``mean field" reduction :
\bea
{\cal G}(\vec n_1\dots \vec n_N)
\to
\prod_i \langle 0 | W^\dagger(0)
{\cal E}(\vec n_i)  W(0) |0\rangle
\eea
reproduces the shift of Eq.\ (\ref{shift}),
because all the shape functions reduce to
the form,
\be
f_e(\epsilon) \to \delta(\epsilon - \lambda_e)\, .
\ee
In this limit, we
  recover the universality of Eq.\ (\ref{lambda}).

The  phenomenology of shape functions has
been discussed in
\cite{Gar01} and in \cite{Taf01},
while in \cite{Bel01},
it was shown that the double distribution
determined phenomenologically in \cite{Taf01},
\bea
f(\varepsilon_L,\varepsilon_R)
&=&
{\rm const}\, (\varepsilon_L\varepsilon_R)^{a-1}
\nonumber\\
&\ & \hspace{-10mm} \times
\exp\left( 
-{\varepsilon_L^2+\varepsilon_R^2+2b\varepsilon_L\varepsilon_R \over 
\Lambda^2}   \right)\, ,
\eea
follows from rather general considerations.
In particular, it was found that the
parameters in this expression have the
interpretations: for $a$, the  number of particles
radiated per unity
rapidity, and for $b$, the correlation due to
a ``spill-over" of radiation at the boundary of the
two hemispheres from hadronization.

\section{CONCLUSION}

The finiteness of time limited the interesting
topics that I was able to cover at Trento,
and made it necessary to pass over developments 
in factorization for exclusive B decay amplitudes \cite{Che01},
in high parton density \cite{Venu}, in hadronic spin structure, and 
in uncertainties
in parton distribution functions \cite{Pum01}.
Many topics have gone unmentioned, but
I hope that I have communicated part of
the reason why many of us find perturbative
QCD and its extensions of enduring interest.

It seems clear to me  that there is much still
to be learned by studying nonperturbative corrections
for observables like those discussed in
this talk.  In some ways, event
shapes related to jets are the ideal testing
ground for the perturbative-nonperturbative
interface.  Perturbation theory is the dominant
contribution, but nonperturbative effects are
substantial and, more importantly, by varying
the energy, and studying moments of distributions, and correlations
of the kind discussed above, we should be able to
``dial" the nonperturbative component.  In this
fashion, we may probe the evolution
of this most demanding of field
theories in new ways, 
providing insight not only into QCD, but into the enterprise
of quantum field theory itself.

\subsection*{Acknowledgements}

My thanks go to the organizers 
of {\it Light Cone Physics: Particles
and Strings}, for their invitation, and to the European
Centre for Theoretical Studies in Nuclear Physics
and Related Areas for its hospitality.  I would especially
like to thank Antonio
Bassetto, Ines Campo, Renzo Leonardi and Federica Vian, for
their generous help. This work was supported in part by the National Science
Foundation, grants PHY9722101 and PHY0098527.

\end{document}